\documentclass{aa}  
\newcommand{\irs}{CrA~IRS\,5}

\usepackage{graphicx}
\usepackage[varg]{txfonts}
\begin{document}
   \title{A Very Long Baseline Interferometry Detection of the Class I Protostar IRS\,5 in Corona Australis}


   \author{Adam T. Deller\inst{1}, Jan Forbrich\inst{2,3}, \and Laurent Loinard\inst{4,5}}

   \institute{The Netherlands Institute for Radio Astronomy (ASTRON), 7990-AA Dwingeloo, The Netherlands
\and
University of Vienna, Department of Astrophysics,         T\"urkenschanzstra{\ss}e 17, 1180 Vienna, Austria
 \and
     Harvard-Smithsonian Center for Astrophysics, 60 Garden Street, Cambridge, MA 02138, USA
\and
Centro de Radiostronom\'{\i}a y Astrof\'{\i}sica, Universidad Nacional Aut\'onoma de M\'exico, 58089 Morelia, Michoac\'an, M\'exico
\and
Max-Planck-Institut f\"ur Radioastronomie, Auf dem H\"ugel 69, D-53121 Bonn, Germany
}

   \date{Received November 30, 2012; accepted -- {\it draft: \today ~JF}}

   \titlerunning{A VLBI Detection of the Class I Protostar CrA IRS\,5}
   \authorrunning{A. T. Deller, J. Forbrich, \& L. Loinard} 


  \abstract
   {}
   {Very Long Baseline Interferometry yields physical constraints on the compact radio emission of young stellar objects. At the same time, such measurements can be used for precise astrometric measurements of parallaxes and proper motions. Here, we aimed to make the first detections of very compact radio emission from class~I protostars in the Corona Australis star--forming region.}
   {We have used the Long Baseline Array (LBA) to observe the protostars IRS~5 and IRS~7 in the Corona Australis star--forming region in three separate epochs.}
   {We report the first firm radio detection of a class~I protostar using very long baseline interferometry. We have detected the previously known non-thermal source IRS~5b, part of a binary system. IRS~5a and IRS~5N were undetected, as were all sources in the IRS~7 region.}
   {These results underline the unusual nature of IRS~5b as a genuine protostar with confirmed non-thermal radio emission. Also, these observations highlight the potential of the LBA as a tool to provide precision astrometric measurements of individual young stellar objects in southern star-forming regions that are not accessible to the Very Long Baseline Array in the northern hemisphere.}


   \maketitle
%

\section{Introduction}

\subsection{Non-thermal radio emission from young stellar objects} 
\enlargethispage{2ex}

Young stellar objects (YSOs) are known to be radio emitters in essentially all evolutionary stages, usually grouped into classes 0--III. The earliest stages appear as class 0/I protostars, followed by disk-dominated class II objects (classical T Tauri stars), with essentially diskless class III sources (also known as weak-line T Tauri stars) representing the final pre-main sequence stage \citep{lad87,and93}. In the past, it has not always been easy to disentangle thermal and non-thermal components to this emission. While many YSOs show thermal emission from ionized material in jets or at the base of outflows, only very few are confirmed non-thermal sources. An unambiguous sign of non-thermal emission is the detection of circular polarization from gyrosynchrotron or linear polarization from synchrotron radiation. A direct determination of high brightness temperatures is an unambiguous sign of non-thermal emission as well. Strong variability is a hint of non-thermal emission, as are negative spectral indices ($\alpha<-0.1$, where $S_\nu \propto \nu^\alpha$; \citealp{andre96a}), but their interpretation is less straightforward \citep{gue02}. 

Non-thermal emission has been detected toward only a handful of YSOs. At radio wavelengths, circular polarization toward T Tauri stars was first found by \citet{whi92cp} and linear polarization was found by \citet{phi96lp}. In earlier evolutionary stages, the first detection of circularly polarized emission toward a class~I protostar was reported by \citet{fei98}. The target was CrA IRS\,5, and this source is essentially still unique in its properties today. The low number of known non-thermal sources among YSOs could be due to the fact that magnetospheric emission is expected to be easily concealed by free-free absorption by ionized material along the line of sight \citep{andre87a}. 

Very long baseline interferometry (VLBI) radio observations yield direct determinations or limits on the brightness temperature of YSOs and thus constitute an important tool in this field. While an obvious target, CrA IRS\,5 is too far south ($\delta \simeq -37$\degr) to be easily observable with the NRAO Very Long Baseline Array. The most prominent YSO studied with VLBI techniques so far is the T Tau system, particularly its southern component which is itself a multiple system and where non-thermal radio emission is clearly detected \citep{phi93,smi03,joh04,loi07a,loi07b} from one of the components (T Tau Sb). However, detailed infrared observations \citep{duch05} have shown that the radio emitting object is most likely a very embedded classical T Tauri star rather than a protostellar source. A more robust case is that of the proto-Herbig Ae/Be object EC95 in the Serpens core \citep{prei99,dzib10}. There, non-thermal radio emission was clearly detected in Very Long Baseline Array (VLBA) observations from the two components of a tight binary system \citep{dzib10}. However, while it is clearly a very young stellar system, EC95 is of intermediate mass so classifying it as Class I is somewhat improper. In an attempt to find more such sources, \citet{for07_vlbi} marginally detected the Ophiuchus source YLW\,15 in VLBI observations, but higher signal-to-noise observations would be needed to confirm this detection.

\subsection{Our target: CrA IRS~5}

IRS\,5 in Corona Australis was first reported when \citet{tay84} discovered the embedded \textit{Coronet} cluster close to the Herbig Ae star R~CrA in near-infrared observations. The distance to this region is $\sim$130~pc (\citealt{cas98}; for a further discussion, see \citealp{neu08}). IRS~5 has been found to be a close binary by \citet{che93}, with a separation of $\sim$0$\farcs$6 or $\sim$78~AU in Very Large Telescope imaging obtained by \citet{nis05}. The latter authors also reported a spectral type of K5--7V for the brighter component, IRS~5a. Finally, a third component of the system (IRS~5N) was found in radio observations \citep{for06,for07}. It is located 8$\farcs$5 north-east of IRS~5a/b, a projected linear separation of $\sim$1100~AU. Subsequent \textit{Spitzer} and Submillimeter Array observations \citep{pet11} have confirmed the classification of the (unresolved) source IRS~5 as a class~I protostar while providing a first classification of IRS~5N, also as a class~I protostar. Interestingly, only IRS~5N and not IRS~5a/b is detected in millimeter continuum observations with the Submillimeter Array.

In the centimeter radio regime, the \textit{Coronet} cluster was first observed by \citet{bro87} who detected radio counterparts to most embedded infrared sources. A multi-epoch radio study of the cluster by \citet{sut96} found that the radio flux density of IRS~5 is variable by a factor of 5 on timescales of weeks, pointing to possible non-thermal emission. Subsequently, \citet{fei98} showed that among the radio counterparts in this cluster, IRS~5 stands out because it shows circular polarization, a clear sign of non-thermal emission. As such, \irs\ has remained one of only very few comparable protostars with clear signs of non-thermal emission. Its non-thermal nature was confirmed by subsequent studies \citep{for06,for07,mie08}. In the deep X-ray study of \citet{fop07}, IRS~5 appears as a marginally resolved source which could be resolved into two counterparts \citep{fpm07conf}. It has since been shown that IRS~5a and 5b have both X-ray and radio counterparts \citep{ham08,cho08}. In a polarimetric study of the region, \citet{cho09} have shown that the non-thermal emission of IRS~5 is entirely due to component IRS~5b, the northeastern source. Since the early classification of the IRS~5 system as a class~I protostar still holds when incorporating \textit{Spitzer} data \citep{pet11}, it is still worth pointing out that, intriguingly, CrA IRS\,5 is essentially the \textit{only} known class~I protostar with clearly non-thermal radio emission. The fact that it is a close binary may offer an explanation since this may help clearing the line of sight of ionized material.

\begin{figure*}
  \centerline{\includegraphics[height=0.95\textwidth,angle=-90]{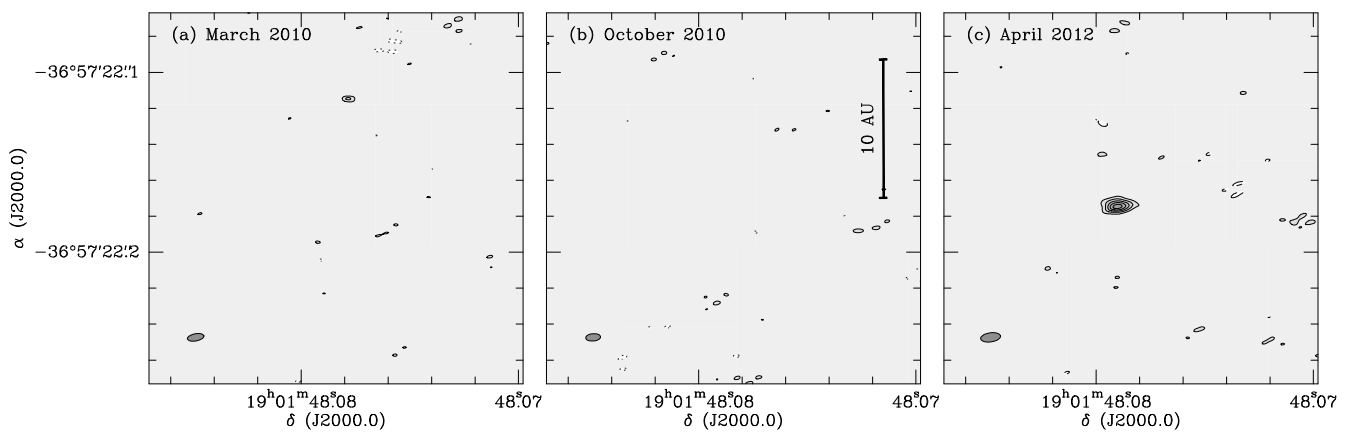}}
  \caption{Naturally weighted 3.6 cm LBA images of \irs\ at the three observed epochs. The synthesised beam is shown in the lower left corner of each image.  The contours show $-$3, $+$3, 5, 7, 9, and 11 $\sigma$, where $\sigma$ is the noise level in each image (65, 75, and 27 $\mu$Jy in the three images, respectively.)  The scale bar in AU assumes a distance of 130 pc.}
    \label{fig:lba}
\end{figure*}

\section{Observations and Data Reduction}

We observed the Coronet region with the Long Baseline Array (LBA) on 3 occasions -- March 2010, October 2010 and April 2012 (LBA project code v329).  The initial goal of these observations was an astrometric campaign to determine the distance to the Coronet region, and accordingly several non-thermal sources  thought likely to be compact (IRS~5 and IRS~7) were targeted in the first two epochs.  Initial positions for IRS~5 and IRS~7 were taken from \textit{Chandra} observations \citep{fop07}.  In the third epoch, only IRS~5 was targeted, for reasons explained below.

In each epoch, continuum observations at 8.4 GHz (3.6 cm) were made using all available antennas of the LBA.  In all 3 epochs, the Australia Telescope National Facility (ATNF) observatories at Parkes (64-m single dish), the Australia Telescope Compact Array (ATCA; five 22-m antennas phased), and Mopra (22-m single dish), were available, as were the Hobart (26-m single dish) and Ceduna (30-m single dish) antennas operated by the University of Tasmania.  For the final observation, the 12-m antenna at Warkworth in New Zealand operated by Auckland University of Technology was also used.  In each case, the observation duration was 10 hours.  Two dual--polarization bands of width 16 MHz were sampled with two bit precision in the frequency ranges 8.409 -- 8.425 GHz and 8.425 -- 8.441 GHz in all observations.  For the final observation, additional bands spanning 8.441 -- 8.457 GHz and 8.457 -- 8.473 GHz were utilized at  the three ATNF stations only (making use of an additional digital backend available at these stations). The data were correlated at Curtin University using the DiFX software correlator \citep{del11} in full polarization mode (correlating both the parallel--hand and cross--hand products).
 
The target sources were phase referenced using a calibrator separated by 2$\degr$ (J1853--3628).  Additionally, a bright fringe finder (1718--649 or 1921--293) was observed for 5 minutes before and after every $\sim$4-hour block on the target field.  For the first two epochs, a cycle time of 6 minutes was used, in which 2 minutes was spent on the calibrator and two minutes on each target source; the predicted 1$\sigma$ image rms for each target field using natural weighting was 63 $\mu$Jy~beam$^{-1}$.  For the final epoch, the cycle time was reduced to 4.25 minutes, of which 3 minutes was spent on the target and 1.25 minutes on the calibrator.  Along with the extra bandwidth, this extra time on a single target field almost doubled the final sensitivity, with a predicted 1$\sigma$ rms of 33 $\mu$Jy~beam$^{-1}$.

Data reduction was performed using the AIPS\footnote{http://www.aips.nrao.edu/index.shtml} software package, and imaging was performed using Difmap \citep{she97}.  The data were amplitude calibrated using system temperature data (logged from the station where available, otherwise a priori values used), followed by delay calibration using the phase reference source. Phase and amplitude calibration were then performed using a model of J1853--3628 derived from these observations.  The absolute amplitude calibration of the data is expected to be correct at the 20\% level -- the low accuracy being due to the limited system temperature logging.  Instrumental polarization was not corrected, as these observations were not intended for high--accuracy polarimetry and did not include scans on a suitable polarization calibrator.  Polarization leakage at the LBA antennas may be as severe as 20\%, although it is likely much lower \citep{dod08a}.

After calibration, we searched for emission over a wide area for each target field using IMAGR in AIPS with 4\,096x4\,096 pixels of size 0.75 milliarcseconds (mas) to cover an area of radius $1\farcs5$.  Fields were imaged at the position of IRS~5a/b, IRS~5N and IRS~7.  For IRS~5a/b, the only field in which a detection was made, the data were corrected to re--center the image on the detection, split, averaged in frequency and exported from AIPS in uvfits format for imaging in Difmap.  In Difmap, $512 \times 512$ pixel images with a pixel size of 0.5 mas were formed using natural weighting for all three epochs.

\section{Results}

\begin{figure}
  \centerline{\includegraphics[width=0.3\textwidth]{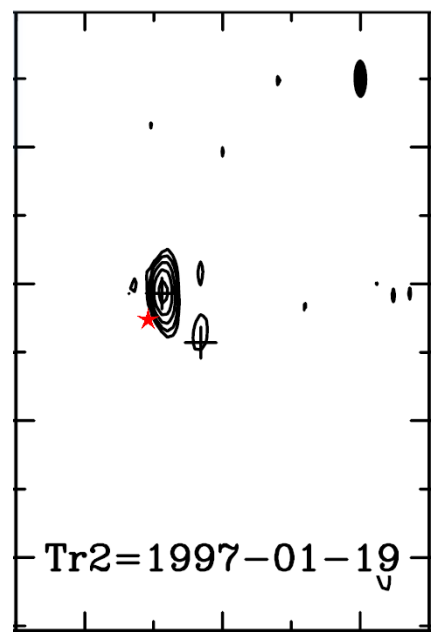}}
  \caption{Highest-resolution (frequency 8.5 GHz, synthesized beam 0$\farcs$7$\times$0\farcs31) VLA image from Fig.~12 in \citet{cho08}.  The tickmarks are spaced at 1\arcsec\ intervals, corresponding to 130 pc.  The position of the source that we detect in the third epoch of our LBA observations is marked by a red star. We interpret the detection as a counterpart of IRS\,5b since IRS\,5a is more than twice as distant. This conclusion is further strengthened when the effects of proper motion are considered.}
    \label{fig:choi08}
\end{figure}

Figure \ref{fig:lba} shows the final IRS~5a/b images from difmap for all three epochs in Stokes $I$.  There is a clear (11$\sigma$) detection in the final epoch with a peak flux density of 330 $\mu$Jy~beam$^{-1}$ at a position of 19$^{\mathrm h}$01$^{\mathrm m}$48$\fs$07899 $-$36\degr57'22$\farcs$1743.  A tentative detection (5.4$\sigma$, discussed further below) with a peak flux density of 350 $\mu$Jy~beam$^{-1}$ is made in the first epoch at a position of 19$^{\mathrm h}$01$^{\mathrm m}$48$\fs$07778 $-$36\degr57'22$\farcs$1150, while no detection is made in the second epoch (5$\sigma$ upper limit of 375 $\mu$Jy~beam$^{-1}$).  As shown in Figure~\ref{fig:choi08} and discussed further below, the source we detect in our LBA images is much closer to IRS~5b than IRS~5a.  Coupled with the previous work summarized in the introduction which identified IRS~5b as a non-thermal source, this positional coincidence allows us to conclude that we have detected IRS~5b, and that IRS~5a was not detected in any epochs.

For the clear detection of IRS~5b in the final epoch, the angular resolution is 5.1~mas $\times$ 11.0~mas, at a position angle of $-$81$\fdg$6.  A single gaussian component was fit to the source using the AIPS task JMFIT -- the fitted deconvolved size is 9.8 $\times$ 5.0 mas at a position angle of 85\degr, with an integrated intensity of 575 $\mu$Jy.  This corresponds to a physical size of 1.3 $\times$ 0.7~AU at the distance of IRS~5.  However, this fit is likely an upper limit to the source size, as residual phase errors (no self-calibration was possible due to the low flux density) are likely to broaden the image somewhat.  No significant signal is seen in Stokes $V$; the level of polarization ($\lesssim$20\%; \citealt{fei98}) seen in earlier observations would make such a detection unlikely given our noise level of 28 $\mu$Jy~beam$^{-1}$ in Stokes $V$ and the substantial instrumental polarization leakage.

We also searched for emission from IRS~5N, using the nominal VLA position from \citet{for06}.  In the final, most sensitive observation, the rms noise level in our LBA images at this position is 28~$\mu$Jy~beam$^{-1}$, similar to the level reached at the position of IRS~5a/b. No potential sources are seen above 5$\sigma$ within $1\farcs5$ of the nominal position, and so we can constrain the peak flux density of the IRS~5N source to be $<$140 $\mu$Jy~beam$^{-1}$ at VLBI scales for this epoch.

Our other target in the \textit{Coronet} cluster, the IRS~7 region, also remained undetected in the two epochs it was targeted, with a 5$\sigma$ upper limit of $\sim$350 $\mu$Jy.  While neither \citet{sut96} nor \citet{for06,for07} had previously found unambiguous signs of non-thermal emission toward protostars in this region, \citet{mie08} reported the tentative detection of gyrosynchrotron radiation from outflow lobes in this area and \citet{cho09} reported the detection of a low degree of circular polarization of IRS~7A in one out of 13 epochs of VLA data analyzed. Our non-detection could thus be due to variability.

\section{Discussion}

The detection obtained here with the LBA of IRS~5b is the first firm VLBI detection of a well-established class~I protostar. Even though both components of this binary were detected in radio emission in earlier observations with lower angular resolution (using the VLA), we only detect one component, IRS 5b. We note that the nearby component IRS 5N also remains undetected.  Based on the peak flux density seen in our VLBI observations, we can derive a lower limit of $>$1.4 $\times$ 10$^5$ K for the brightness temperature of IRS~5b.  This lower limit is too low to rule out thermal emission by itself and thus does not provide a clear signature of non-thermal emission. Such a signature was, however, provided earlier with multiple reported detections of circularly polarized emission from this source (see above).  The flux density at which the source is detected is compatible with the quiescent flux densities found by \citet{for06} and \citet{cho08}.
These observations constrain the size of the emission region to less than 1~AU at the distance of CrA, by far the best constraint provided by direct imaging to date.  However, this is still much larger than the presumed size of this low-mass star, and the non-thermal emission is most likely due to coronal activity which extends at most out to a small number of stellar radii.

\begin{figure}
\centerline{
\begin{tabular}{c}
\includegraphics[width=0.45\textwidth]{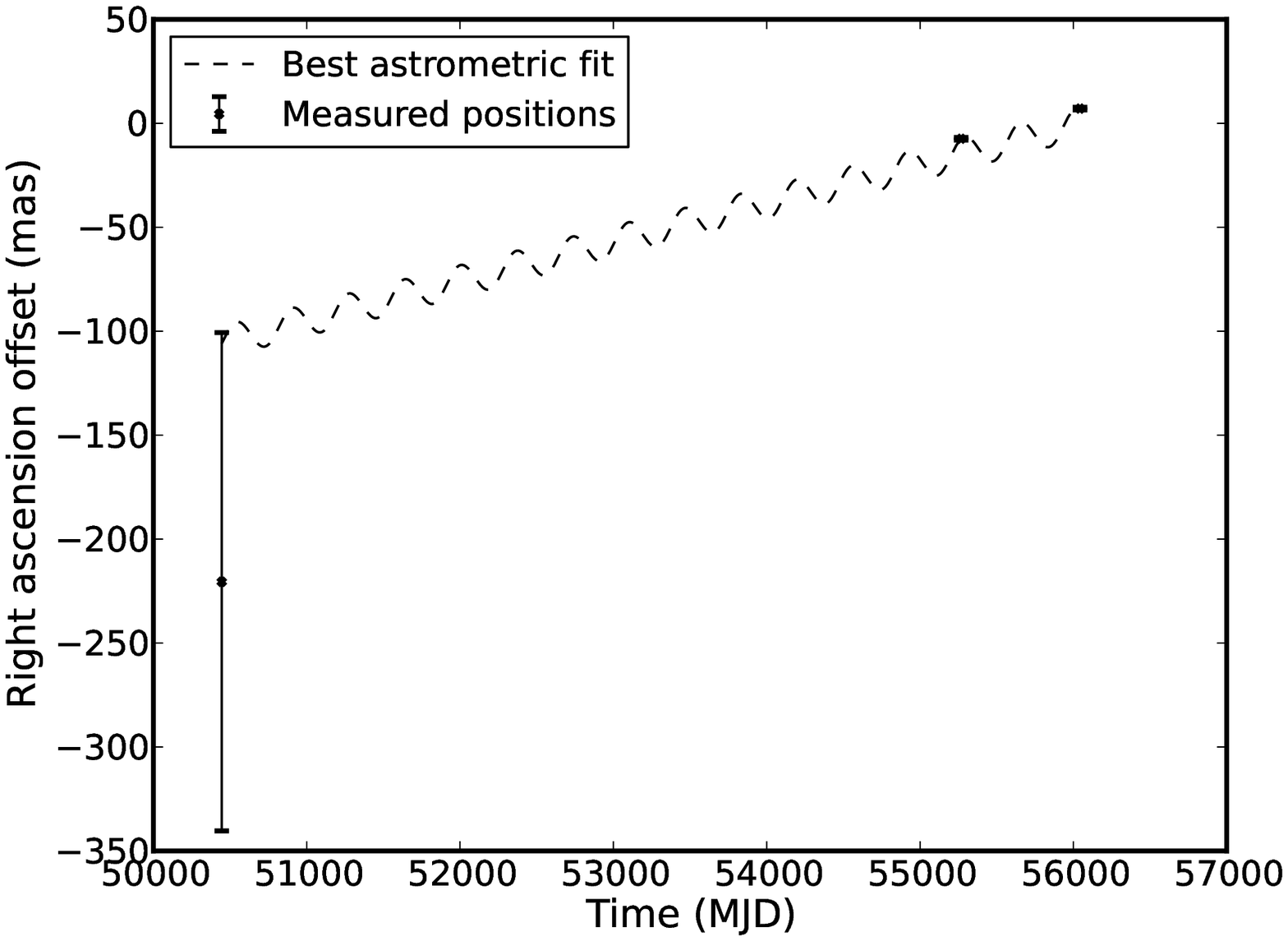} \\
\includegraphics[width=0.45\textwidth]{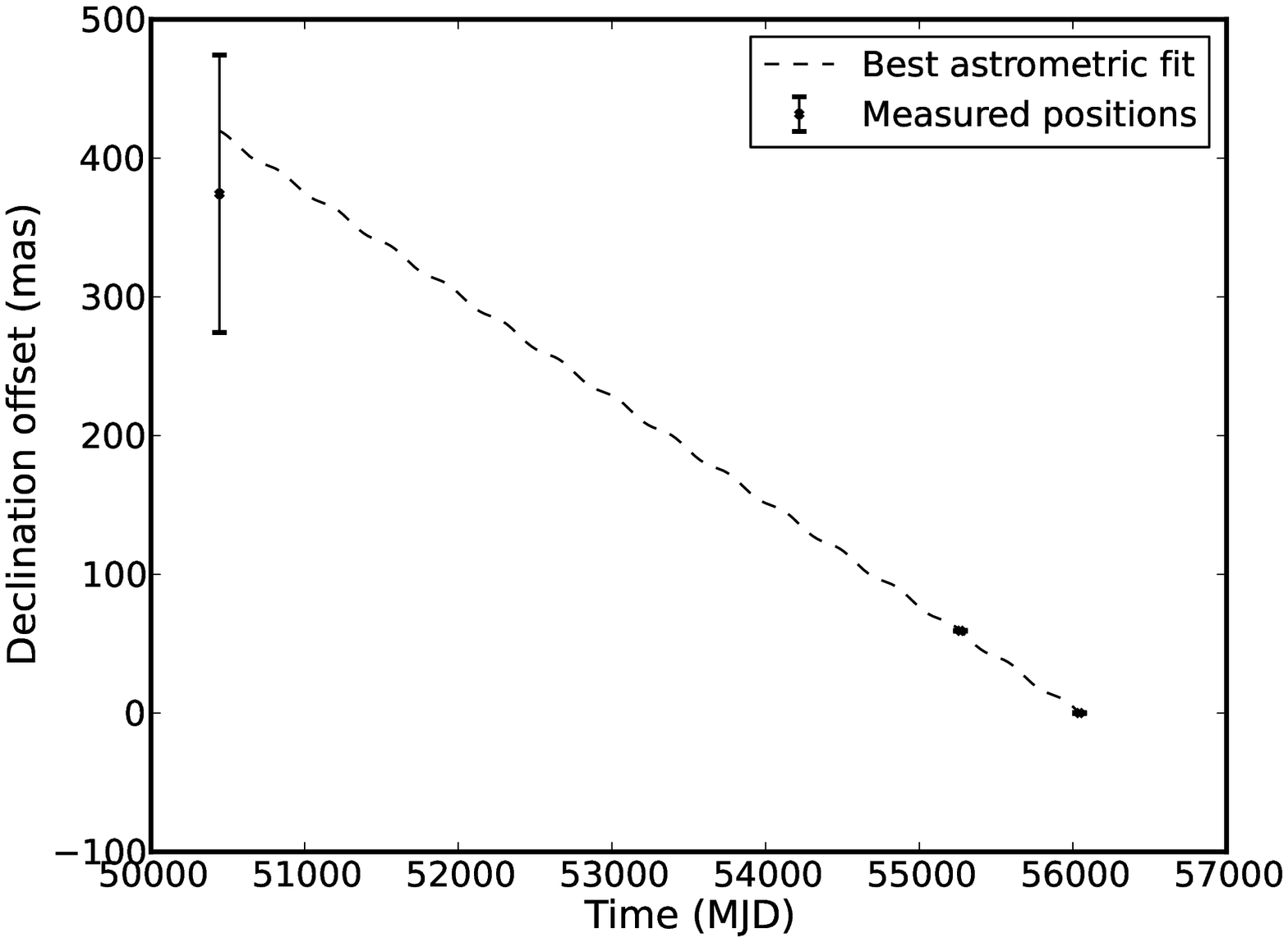}
\end{tabular}
}
\caption{Proper motion of IRS~5b fitted from the two LBA detections and one VLA position taken from \cite{cho08}.  The parallax has been fixed based on the assumed distance of 130 pc, but the proper motion results are insensitive to small variations in this parameter.}
\label{fig:pmfit}
\end{figure}

As noted above, we consider the 5.4$\sigma$ peak in the Stokes $I$ image of IRS~5 from the first epoch observation to be a tentative detection, an interpretation which is strengthened by the proper motion which it implies.  In addition to the two position measurements of IRS~5b from our first and third observing epochs, we can make use of Very Large Array (VLA) observations presented by \citet{cho08} to greatly extend the time baseline of these radio observations and check for consistency.  We use the reference position for IRS~5b from the observation Tr 1 in \citet{cho08}, which offers the longest time baseline and the highest angular resolution (VLA A-array observations made in 1996 at a wavelength of 3.5 cm).  With these three points, we are able to fit for a radio proper motion, which we find to be $\mu_\alpha$cos$\delta$ = 6.9$\pm$0.3 mas yr$^{-1}$, $\mu_\delta$ = -27.3$\pm$0.1 mas yr$^{-1}$.  This fit is dominated by the two LBA positions, since the error bars are much smaller, but the fit is consistent with all 3 epochs.  Furthermore,  it closely matches the value of  ($\mu_\alpha$cos$\delta$, $\mu_\delta$) = (5.5, $-$27.0)\,mas\,yr$^{-1}$ obtained for the mean proper motion of CrA members  by \citet{neu00}. To perform this fit, the parallax was fixed to 7.7 mas, representing a distance of 130 pc.  The positions and fitted motion of IRS~5b are shown in Figure~\ref{fig:pmfit}.  We note that \citet{cho08} were unable to determine a significant proper motion from VLA observations of IRS~5b alone, but the LBA observations added here extend the time baseline and have a resolution and positional accuracy more than an order of magnitude higher.

The nominal positional accuracy of our significant VLBI detection in the third observing epoch (as reported by the AIPS task JMFIT) is 0.6 mas in right ascension and 0.3 mas in declination.  Given that the separation to the primary calibrator is 2\degr, systematic astrometric errors for LBA observations are unlikely to exceed this level \citep{del08,del12}.  With 4 such epochs, it should be possible to measure the parallax and proper motion of IRS~5b at the level of 0.2 mas and 0.2 mas yr$^{-1}$ respectively, sufficient to obtain the distance to the source with an accuracy of only a few pc.  However, obtaining this result will require significant additional observing resources, as a full observing track at each epoch is required to obtain even a low--significance detection, and the effect of variability is still uncertain, since we have only one firm detection.  Over the relevant observing timescale (several years) the binary orbital motion would be negligible, since the binary separation is $>$100~AU \citep{cho08}.

\section{Summary}
This first VLBI detection of a class~I protostar underlines the peculiar nature of CrA IRS\,5.  Compared to previous detections, IRS 5b shows a relatively low radio flux density in our detections. Future observations, if coinciding with a more active state of the system (like that seen by \citealt{cho08}) would place considerably more stringent constraints on the nature of the emission.  Furthermore, our detection shows the potential for VLBI observations of these sources to determine accurate parallax and proper motion measurements in an absolute reference frame.  Our preliminary results already indicate that the proper motion of IRS~5b is similar to that of other members of the \textit{Coronet} cluster.  The currently best distance determination of CrA is based on a full orbit solution of the double-lined spectroscopic binary TY~CrA \citep{cas98}, yielding a distance of 128$\pm$11~pc. However, that source is not located directly in the \textit{Coronet} cluster, and in the past, distances of up to 170~pc have been estimated for CrA. Our LBA measurements demonstrate that distance determinations to star-forming regions, based on parallax measurements using continuum emission from young stellar objects (e.g., \citealp{men07,loi07b,dzib10}) is also possible at southern declinations that are not accessible with the NRAO VLBA. Given the low detection rate of non-thermal emission in protostars, our observations also show the importance of carefully selecting suitable VLBI targets from preceding measurements at lower angular resolution.

\begin{acknowledgements}
The Long Baseline Array is part of the Australia Telescope National Facility which is funded by the Commonwealth of Australia for operation as a National Facility managed by CSIRO. A.T.D. was supported by an NWO Veni Fellowship. This publication is supported by the Austrian Science Fund (FWF). L.L. acknowledges the financial support of DGAPA, UNAM, CONACyT (M\'exico), and the Alexander von Humboldt Stiftung.  
\end{acknowledgements}

\bibliographystyle{aa} 
\bibliography{bib_irs5vlbi} 

\end{document}